\documentclass[reprint,aps,prb,superscriptaddress,amsmath,amssymb,bibnotes,showkeys]{revtex4-1}
\usepackage{graphicx}

\usepackage{graphicx}
\usepackage{amsmath}
\usepackage{dcolumn}
\usepackage{bm}
\usepackage{braket} 
\usepackage{bbold}
\usepackage{soul}
\usepackage{diagbox}
\usepackage{multirow}
\usepackage{url}
\usepackage{tabularx}
\usepackage{color}
\usepackage[
 colorlinks=true,
 urlcolor=blue,
 citecolor=blue,
 linkcolor=blue,
 bookmarks=false,
 pdfstartview={FitH},
]{hyperref}

\def\cm{cm$^{-1}$}

\begin{document}

\title{Raman scattering study of NaFe$_{0.53}$Cu$_{0.47}$As}

\author{W.-L.~Zhang}
\email{wz131@physics.rutgers.edu}
\affiliation{Department of Physics $\&$ Astronomy, Rutgers
University, Piscataway, New Jersey 08854, USA} 
\author{Y. Song}
\affiliation{Department of Physics and Astronomy and Rice Center for
Quantum Materials, Rice University, Houston, Texas 77005, USA}
\author{W.-Y. Wang}
\affiliation{Department of Physics and Astronomy and Rice Center for
Quantum Materials, Rice University, Houston, Texas 77005, USA}
\author{C.-D. Cao}
\altaffiliation{Current address: Department of Applied Physics, 
Northwestern Polytechnical University, Xi'an 710072, China}
\affiliation{Department of Physics and Astronomy and Rice Center for
Quantum Materials, Rice University, Houston, Texas 77005, USA}
\author{P.-C. Dai}
\affiliation{Department of Physics and Astronomy and Rice Center for
Quantum Materials, Rice University, Houston, Texas 77005, USA}
\author{C.-Q. Jin}
\affiliation{Beijing National Laboratory for Condensed Matter Physics 
and Institute of Physics, Chinese Academy of Sciences, Beijing 
100190, China}  
\affiliation{Collaborative Innovation Center of Quantum Matter, Beijing, China}
\author{G.~Blumberg}
\email{girsh@physics.rutgers.edu}
\affiliation{Department of Physics $\&$ Astronomy, Rutgers
University, Piscataway, New Jersey 08854, USA}
\affiliation{National Institute of Chemical Physics and Biophysics,
Akadeemia tee 23, 12618 Tallinn, Estonia}

\date{\today}

\begin{abstract}

We use polarization-resolved Raman scattering to study lattice
dynamics in NaFe$_{0.53}$Cu$_{0.47}$As single crystals. 
We identify four $A_{1g}$ phonon modes, at 126, 172, 183, and 197 \cm, and
four $B_{3g}$ phonon modes at 101, 139, 173, and 226 \cm ($D_{4h}$ point group). 
The phonon spectra
are consistent with the $Ibam$ space group, which confirms that the Cu and
Fe atoms form a stripe order. 
The temperature dependence of the
phonon spectra suggests weak electron-phonon and magneto-elastic
interactions.

\end{abstract}

\maketitle

\section{introduction}
The parent compound of the iron-pnictide superconductor,  NaFeAs, is a ``bad
metal.'' 
It exhibits a tetragonal-to-orthorhombic transition at
52~K, a paramagnetic-to-spin-density wave transition at
41~K, and a superconducting transition at
23~K~\cite{GFChen_PRL2009}. 
Doping copper into NaFeAs 
suppresses both the orthorhombic and the paramagnetic-to-spin-density-wave orders and enhances the
superconductivity~\cite{Blundell_RPB2012,Fang_PRB2013,Tan_PRB2017}.
Recently, it was shown that heavy Cu substitution at the Fe site induces
Mott-insulator-like behavior~\cite{Wang_PRX2015,Songyu_ncommu2016}.
The electronic properties of the heavily doped NaFe$_{1-x}$Cu$_x$As 
are similar to those of lightly
doped cuprates~\cite{Wang_PRX2015,Shi_PRL2016,Basov_PRB2017}. 

\begin{figure}[t]
\includegraphics[width=0.9\columnwidth]{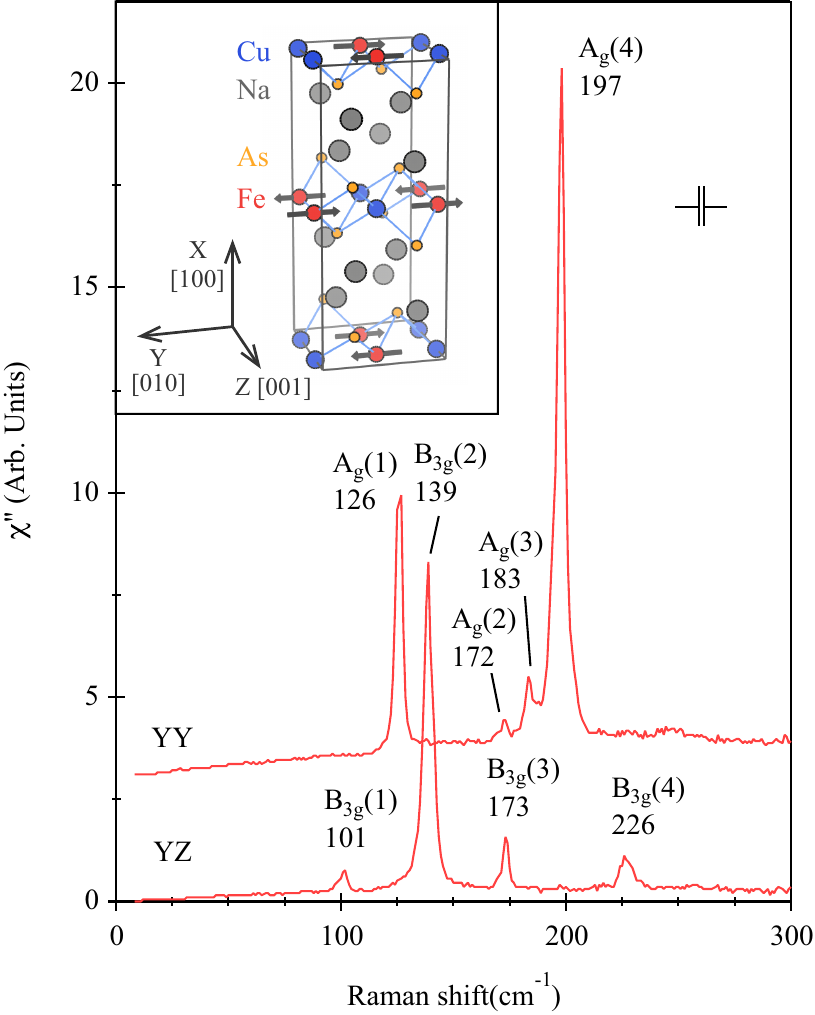}
\caption{\label{Fig1} 
Raman scattering spectra of NaFe$_{0.53}$Cu$_{0.47}$As crystals for
YY + ZZ and YZ + ZY scattering geometries at 250 K measured with 1.9~eV excitation. The spectral resolution is 2.5 cm$^{-1}$.
Inset: NaFe$_{0.5}$Cu$_{0.5}$As unit cell with Cu and Fe
collinear stripe order. 
Arrows at the Fe sites mark magnetic moments.
}
\end{figure}

For Cu substitution concentration $x > 0.44$ a long-range collinear
antiferromagnetic order with magnetic moments residing  
only at the Fe sites develops below 200~K.  
The moment increases with Cu
concentration substitution $x$~\cite{Songyu_ncommu2016}. 
At the solubility limit near $x = 0.5$, new superlattice peaks appear in 
the TEM diffraction pattern, which are interpreted as the signature 
of Cu and Fe stripe order formation~\cite{Songyu_ncommu2016}, as
depicted in the inset in Fig.~\ref{Fig1}.
Compared to the parent NaFeAs
compound in the tetragonal phase, the
stripe-ordering of Cu and Fe in heavily doped NaFe$_{1-x}$Cu$_x$As
removes the lattice four-fold rotational symmetry and reduces the
crystallographic space group from $Fmmm$ (point group $D_{4h}$) to
$Ibam$ (point group $D_{2h}$), making a structural analog of the
magnetic order in parent NaFeAs crystals.

Here we present a polarization-resolved Raman
scattering study of the lattice dynamics for 
NaFe$_{0.53}$Cu$_{0.47}$As single crystals. 
Four $A_{g}$ phonon modes, at 126,
172, 183, and 197~cm$^{-1}$ and four $B_{3g}$ phonon modes, at 101, 139,
173, and 226~cm$^{-1}$, are identified. 
The phonon spectra are consistent
with the Fe/Cu stripe-ordered structure. All the observed phonons
exhibit a symmetric line shape. Across the antiferromagnetic phase transition, no
phonon anomaly is observed. The data suggest weak electron-phonon
and magneto-elastic interaction. 

\section{experimental}

NaFe$_{1-x}$Cu$_{x}$As single crystals were grown by the self-flux
method~\cite{Tanatar_PRB2012,Songyu_ncommu2016}. 
The nominal Cu
concentration was $x$ = 0.85, which resulted in an actual concentration $x$ = 0.47 ~\cite{Songyu_ncommu2016}. 
The preparation of the reference LiFeAs single crystal is described in~\cite{Wang2010}.

\begin{figure}[!t]
\includegraphics[width=0.9\columnwidth]{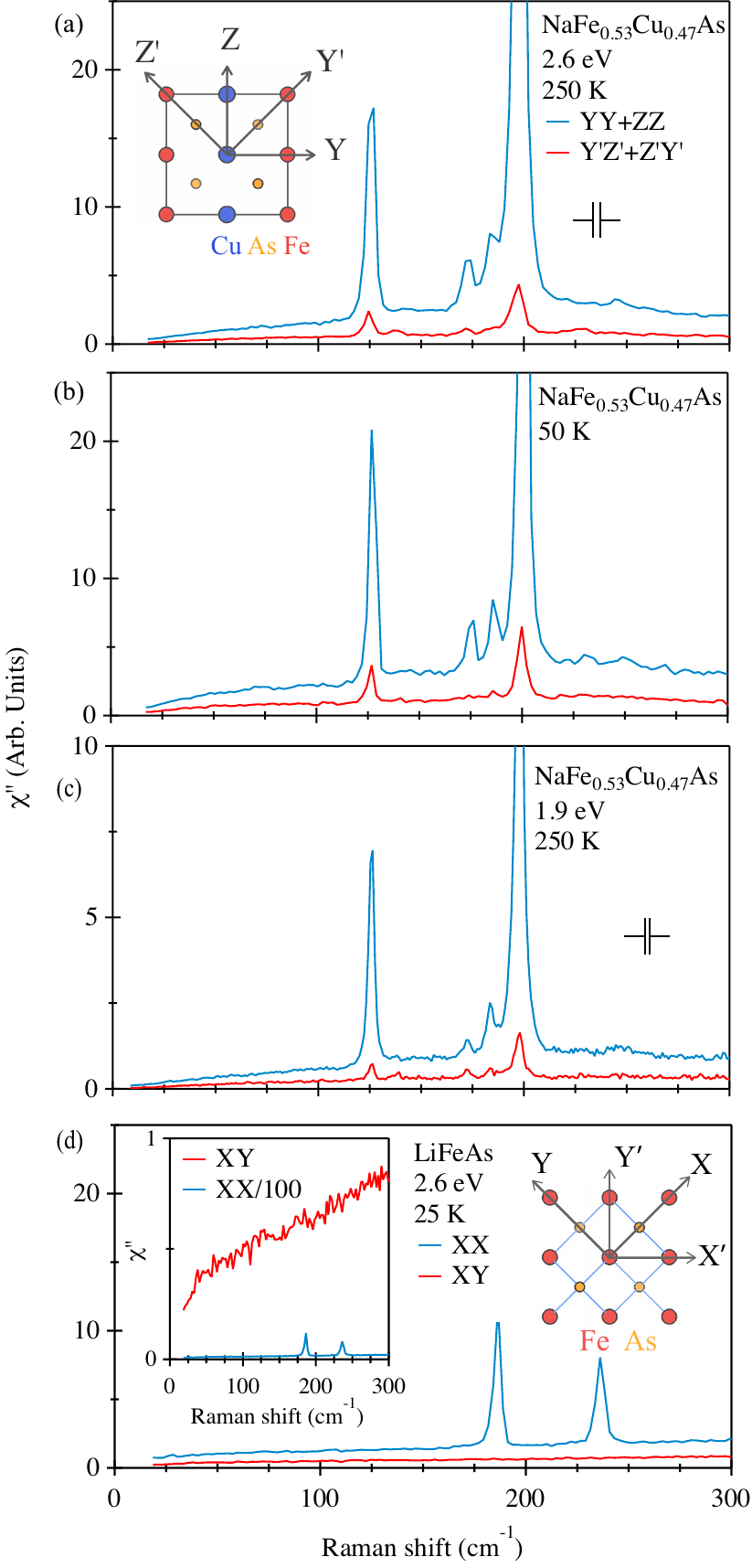}
\caption{\label{Fig2} 
(a, b) $A_g$-symmetry Raman active phonon modes measured for 
NaFe$_{0.53}$Cu$_{0.47}$As crystals at (a) 250 K 
and (b) 50~K in YY + ZZ (blue line) and 
Y$^\prime$Z$^\prime$+Z$^\prime$Y$^\prime$ (red line) 
scattering geometries with 2.6~eV laser excitation with spectral resolution 3.5~cm$^{-1}$.  Inset in (a): top view of the
Fe-Cu-As layer for NaFe$_{0.53}$Cu$_{0.47}$As structure and the 
YZ-Y$^\prime$Z$^\prime$ coordinates.  
(c) Same  $A_g$ phonon modes measured at 250~K with 1.9~eV excitation. 
(d) Raman spectra from tetragonal LiFeAs crystal at 25~K measured in 
X$^\prime$X$^\prime$ (blue) and XY (red) scattering geometries 
with 2.6~eV-laser excitation.  
Inset in (d): 
Left: Zoom in of the data where the signal for X$^\prime$X$^\prime$ polarization is divided 
by 100 to demonstrate the lack of detectable \emph{leakage} into cross polarization.
Right: Top view of the
Fe-As layer for the LiFeAs crystal structure and the 
XY-X$^\prime$Y$^\prime$ coordinates.  
}
\end{figure}

The NaFe$_{1-x}$Cu$_{x}$As crystal belongs to the $Ibam$ space group at room temperature, as shown in the 
inset in Fig.~\ref{Fig1}. 
The crystallographic principal axis [001] of the
$Ibam$ group is along the Fe(Cu) stripe direction. We define the X, Y, and Z
axes along crystallographic [100], [010], and [001] axes and
Y$^\prime$/Z$^\prime$ along the [011]/[0$\bar{1}$1] direction (inset
Fig.~\ref{Fig2}(a)). 
 
There are 12 atoms in the primitive unit cell. 
Group theoretical analysis infers 
$4A_g+6B_{1g}+4B_{2g}+4B_{3g}+2A_u+4B_{1u}+6B_{2u}+6B_{3u}$~\cite{Bilbao_1}
symmetry decomposition of the 36 phonon modes at
the Brillouin center $\Gamma$ point.
All the even $g$ modes are Raman active. The irreducible
representations and decomposition of the Raman active modes by
symmetry are
summarized in Table~\ref{Table1}.
\begin{table}[!t]
\renewcommand\arraystretch{1.2}
\caption{
Phonon mode decomposition at the $\Gamma$ point and selection rules for 
Raman-active modes in $Ibam$ space group.}
\label{Table1}
\centering
\begin{tabularx}{\columnwidth}{lcr}
\hline\hline
&& Irreducible representation\\
Acoustic	&&B$_{1u}$+B$_{2u}$+B$_{3u}$\\
IR 		&&3B$_{1u}$+5B$_{2u}$+5B$_{3u}$\\
Raman	&&4A$_{g}$+6B$_{1g}$+4B$_{2g}$+4B$_{3g}$\\
Silent	&&2A$_u$\\
\hline
Atom &Wyckoff position&Raman active mode\\
Na 		&	8j	&2A$_g$+2B$_{1g}$+B$_{2g}$+B$_{3g}$\\
Fe 		&	4b	&B$_{1g}$+B$_{2g}$+B$_{3g}$\\
Cu 		&	4a	&B$_{1g}$+B$_{2g}$+B$_{3g}$\\
As 		&	8j	&2A$_g$+2B$_{1g}$+B$_{2g}$+B$_{3g}$\\
\hline
\hline
\end{tabularx}
\end{table}%

Polarization-resolved low-temperature Raman scattering measurements 
were performed in a quasi-back 
scattering setup
from natural cleaved (100) surface~\cite{Gozar_dissertation}. 
Polarizers with an extinction ratio better than 1:500 were employed  
\footnote{In the set-up, see Fig. 2.1 in~\cite{Gozar_dissertation}, 
Melles Griot Glan-Taylor polarizing prism with better  
than $10^{-5}$ extinction ratio was used to clean the laser excitation 
beam and Karl Lambrecht Corporation broad band polarizing cube with 
extinction ratio better than 1:500 was used for the analyzer}. 
Samples were cleaved in a nitrogen-filled glove bag and immediately
transferred to an optical cryostat with continuous helium gas flow. 
We used
1.9 and 2.6~eV excitations from a Kr$^+$ laser, where the laser was
focused on a 50$\times$50~$\mu m^2$ spot on the sample. The power was
kept below 10 mW to minimize the laser heating. 
The local laser heating was 
estimated~\cite{WZhang_PRB2016,Wu_1712long} and kept at less than 5~K.  
All referred temperatures are corrected for the laser
heating.

\begin{table}[!b]
\renewcommand\arraystretch{1.2}
\caption{Raman tensor and selection rules for Raman-active modes
in the $D_{2h}$ group.}
\label{Table2}
\centering
\begin{tabularx}{\columnwidth}{ccccc}
\hline\hline

&&&&\\

\multicolumn{2}{c}{$R_{A_{g}} = \left[ \begin{array}{ccc} a & 0 & 0
\\ 0 & b & 0 \\ 0 & 0 & c\end{array}\right]$ }
&\multicolumn{3}{c}{$R_{B_{1g}} =  \left[ \begin{array}{ccc} 0 & d &
0  \\ e & 0 & 0 \\ 0 & 0 & 0\end{array}\right]$}\\
&&&&\\
\multicolumn{2}{c}{$R_{B_{2g}} = \left[ \begin{array}{ccc} 0 & 0 & f
\\ 0 & 0 & 0 \\ g & 0 & 0\end{array}\right]$
}&\multicolumn{3}{c}{$R_{B_{3g}} =  \left[ \begin{array}{ccc} 0 & 0 &
0  \\ 0 &0& h \\ 0 & i & 0\end{array}\right]$}\\
&&&&\\

(001) surface	&XX	&~~~YY	&~~~XY/YX	&
\\
$A_g$ 		&$a^2$		&~~~$b^2$		&~~~0			&	\\
$B_{1g}$ 		&0			&~~~0			&~~~$d^2$/$e^2$	&	\\

(010) surface	&XX&~~~ZZ	&~~~XZ/ZX
&\\
$A_g$ 		&$a^2$		&~~~$c^2$		&~~~0			&	\\
$B_{2g}$ 		&0			&~~~0			&~~~$f^2$/$g^2$	&	\\

(100) surface
&YY/ZZ&~~YZ/ZY&~~~Y$^\prime$Y$^\prime$/Z$^\prime$Z$^\prime$&Y$^\prime$Z$^\prime$/Z$^\prime$Y$^\prime$\\
$A_g$ 		&$b^2$/$c^2$	&~~~0			&~~~$(b+c)^2/4$	&~~~$(b-c)^2/4$\\
$B_{3g}$ 		&0			&~~~$h^2$/$i^2$	&~~~$(h+i)^2/4$	&~~~$(h-i)^2/4$	\\

\hline
\hline
\end{tabularx}

\end{table}%

The Raman scattering signal was analyzed with a triple-stage
spectrometer with the spectral resolution setting at about 2 \cm. 
We used scattering geometries $\mu\nu$ with $\mu$/$\nu$ = Y, Z,
Y$^\prime$ and
Z$^\prime$, where $\mu\nu$ is short for $\bar{X}$($\mu\nu$)X in
Porto's notation.  All spectra were
corrected for the spectral response to obtain the Raman scattering
intensity $I_{\mu\nu}(\omega,T)$. The Raman susceptibility
$\chi_{\mu\nu}^{\prime\prime}(\omega,T)$ was related to
$I_{\mu\nu}(\omega,T)$ by $I_{\mu\nu}(\omega,T)$ =
$\chi_{\mu\nu}^{\prime\prime}(\omega,T)[1+n(\omega, T)]$, where
$n(\omega, T)$ is the Bose factor.

In Table~\ref{Table2} we list the Raman tensor for the $D_{2h}$ group and
the selection rule for experimentally accessible
polarizations~\cite{Cardona_book1982}. 
Due to the twin structure~\cite{Songyu_ncommu2016}, the
collected signal from the (100) surface is a superposition of Raman
scattering intensities from two types of orthogonal
domains. 
For example, the signal for parallel polarized scattering geometry 
along the crystallographic axes contains the intensity from YY 
geometry for one type 
of domain and ZZ geometry for the other type of domain. 
We denote this scattering geometry as YY + ZZ. 
Similarly, cross polarized signal 
along the crystallographic axes contains contributions from 
YZ and ZY geometries and is denoted YZ + ZY, 
and cross polarized signal along the diagonal directions contains 
contributions from Y$^\prime$Z$^\prime$ and Z$^\prime$Y$^\prime$ 
scattering geometries, is denoted 
Y$^\prime$Z$^\prime$ + Z$^\prime$Y$^\prime$.

Following the notation in Table~\ref{Table2}, we assign all phonons that appear in
the YY + ZZ geometry to the $A_{g}$ symmetry modes, and those appear in
the YZ + ZY geometry to the $B_{3g}$ modes. 

%
%
%
%
%
%
\begin{figure}[t]
\includegraphics[width=\columnwidth]{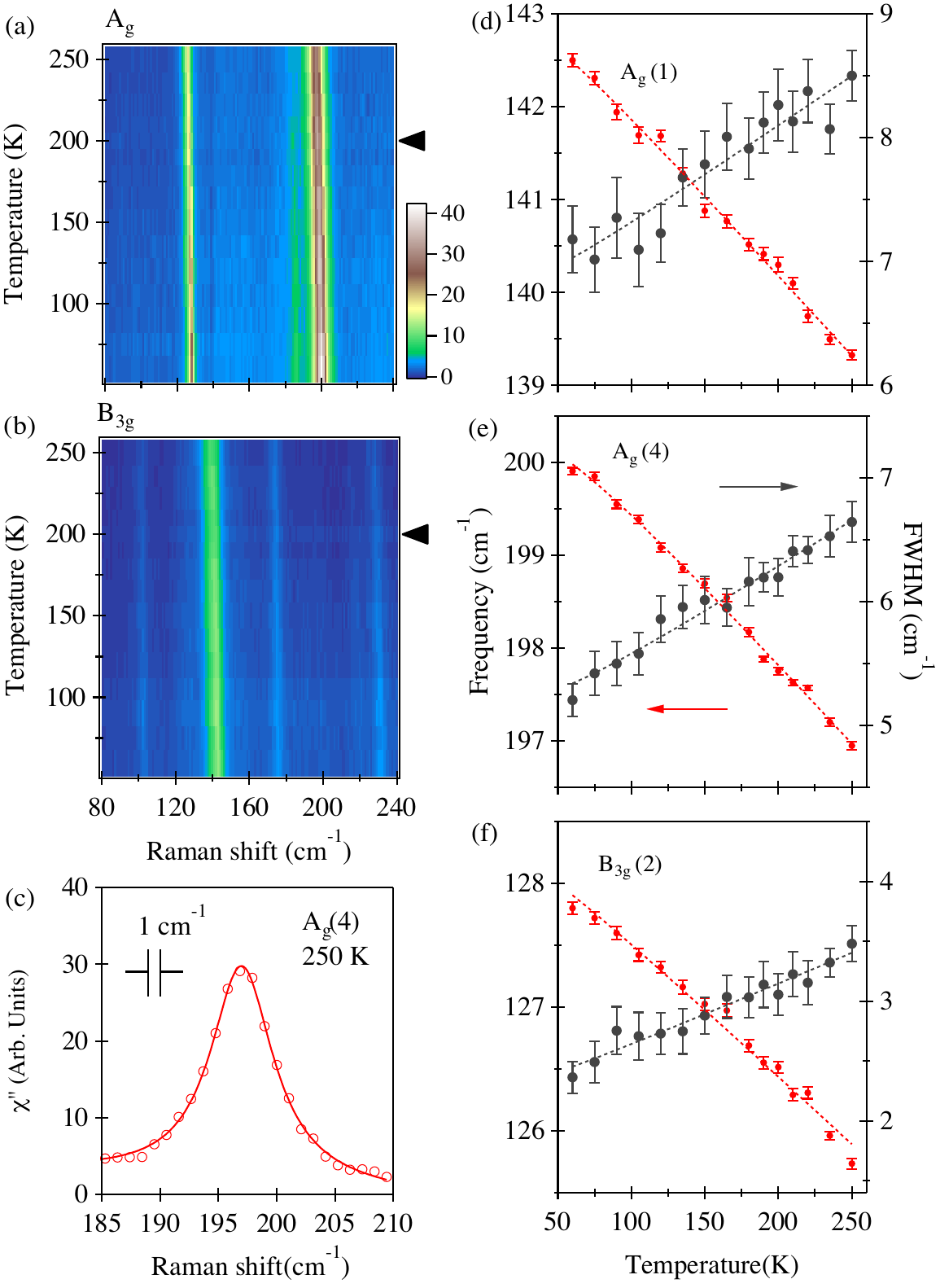}
\caption{\label{Fig3} 
For NaFe$_{0.53}$Cu$_{0.47}$As crystals, temperature dependence of the 
Raman response in (a) $A_g$ and (b)  
$B_{3g}$ symmetry channels 
measured with 1.9~eV laser excitation. 
The spectral resolution is 1~cm$^{-1}$. 
Black arrows indicate the magnetic phase transition at 200 K. 
(c) Lorentz fit to the $A_g(4)$ phonon at 250 K. 
Inset: Spectral resolution. 
(d-f) Temperature dependence of the phonon peak frequency for the 
$A_g(1)$, $A_g(4)$, and $B_{3g}$(2) modes.  
Vertical error bars are one standard deviation error of the Lorentzian fit. 
Dashed lines show fits of the phonon frequency and line width to 
Eqs.~(\ref{Eq1}) and (\ref{Eq2}).}
\end{figure}
%
%

\section{Results and Discussion}

In Fig.~\ref{Fig1} we show the Raman response for
NaFe$_{0.53}$Cu$_{0.47}$As crystals at 250~K for YY + ZZ and YZ + ZY scattering
geometries. 
We identify all the $A_{g}$ and $B_{3g}$ phonon modes predicted
by group theory: 
four $A_{g}$ symmetry modes, at 126, 172,
183, and 197~cm$^{-1}$, and four $B_{3g}$ symmetry modes, at 101,
139, 173, and 226~cm$^{-1}$. 
All modes show a symmetric line shape.

We note that at the same frequency as the $A_g$ phonon modes, some 
modes with a weaker intensity are also observed for the
Y$^\prime$Z$^\prime$ + Z$^\prime$Y$^\prime$ geometry for both 2.6 
and 1.9~eV laser excitations [Figs.~\ref{Fig2}(a)-(c)].   
The intensity of the \emph{leaking} modes is
about 10\% of the $A_g$ phonon intensity in the YY+ZZ geometry, which 
is much higher than 
the experimental polarization extinction ratio.
In Fig.~\ref{Fig2}(d) we show data for the LiFeAs tetragonal structure~\cite{Tacon_PRB2012} 
measured employing the same setup.
If the substituted Cu ions at Fe sites were randomly disordered, 
the NaFe$_{1-x}$Cu$_x$As structure would have the same 
point-group symmetry as the LiFeAs structure.
By symmetry, no Raman-active phonons are allowed in the XY scattering 
geometry for the LiFeAs structure. 
As we demonstrate in the inset in Fig.~\ref{Fig2}(d), the  
\emph{leakage} intensity for the tetragonal LiFeAs structure is less than 
a percent.

Based on the Raman scattering selection rules, we can deduce that the 
\emph{leakage} intensity is proportional to $(b-c)^2/4$ 
(Table~\ref{Table1}), which is a measure of the anisotropic 
electronic properties between the Y and the Z directions~\cite{Wu_1712long}.  
The observation of the \emph{leakage} 
is consistent with the suggested
formation of a long range stripe order 
which breaks the crystallographic four-fold symmetry~\cite{Songyu_ncommu2016}. 
The count of observed Raman-active phonons for the NaFe$_{1-x}$Cu$_x$As 
structure also suggests that the size of its primitive cell is  four 
times larger than that for the NaFeAs structure (Table~\ref{Table1}), 
therefore, the only possible consistent structure is the Fe-Cu stripe 
order phase, as shown in the inset in Fig.~\ref{Fig1}.

In Figs.~\ref{Fig3}(a) and (b) we show the intensity plot of the Raman 
response $\chi^{\prime\prime}(\omega,T)$ for $A_g$ (YY + ZZ) and 
$B_{3g}$ (YZ + ZY) symmetry channels between 250 and 60~K.  
All phonons show a symmetric line shape. The number of phonon modes 
and their line shapes do not change across the antiferromagnetic phase transition 
at 200~K, suggesting weak magneto-elastic interaction.

We analyze $A_g(1)$, $A_g(4)$ and $B_{3g}(2)$ phonons by fitting to 
Lorentzian function. 
As an example, Fig.~\ref{Fig3}(c) shows the $A_g(4)$ mode at 
250~K and its Lorentzian fit. 
The fitting results are summarized in Figs.~\ref{Fig3}(d) and (e). 
Since the magneto-elastic interaction appears to be undetectable 
within the experimental resolution, we fit the modes temperature dependence by       
the anharmonic decay model for the entire temperature range (250 to 
60~K)~\cite{Cardona_PRB1984}:    
\begin{equation}
\label{Eq1}
\omega(T)=\omega_0-\omega_1[1+\frac{2}{e^{\hbar\omega_0/2k_BT}-1}]
\end{equation}
\begin{equation}
\label{Eq2}
\Gamma(T)=\Gamma_0+\Gamma_1[1+\frac{2}{e^{\hbar\omega_0/2k_BT}-1}]
\end{equation}
The fitting results are summarized in Table~\ref{Table3}.

\begin{table}[!t]
\renewcommand\arraystretch{1.4}
\caption{Fitting parameters for the frequency and linewidth of the 
$A_g$(1), $A_g$(4), and $B_{3g}$(2) modes. 
Units are~cm$^{-1}$.}   
\label{Table3}
\centering
\begin{tabularx}{\columnwidth}{c|cccc}
\hline
\hline
~~~Mode ~~~	&~~~~~~$\omega_0$~~~~~~ 		&~~~~~~$\omega_1$~~~~~		&~~~~~~$2\Gamma_0$~~~~		&~~~~~~$2\Gamma_1$~~~~	\\
\hline
$A_g$(1) 	&128.72$\pm$0.07	&0.49$\pm$0.02	&2.06$\pm$0.06	&0.23$\pm$0.02\\
$A_g$(4)	&201.51$\pm$0.06	&1.21$\pm$0.02	&4.67$\pm$0.07	&0.53$\pm$0.03\\
$B_{3g}$(2)&143.82$\pm$0.07	&0.87$\pm$0.02	&6.4$\pm$0.1		&0.41$\pm$0.04\\
\hline
\hline

\end{tabularx}
\end{table}%

\section{conclusion}

In summary, we present a polarization-resolved Raman
scattering study of NaFe$_{0.53}$Cu$_{0.47}$As single
crystals. 
We observe four $A_g$ and  four $B_{3g}$ phonon modes at 126, 172,
183, and 197~cm$^{-1}$ and 101, 139, 173, and 226 cm$^{-1}$, respectively. 
The results are consistent with the $Ibam$ space-group symmetry structure
where Fe/Cu atoms form a stripe order.
No phonon anomaly is observed cross the magnetic phase
transition from 250 to 60 K, suggesting weak electron-phonon and
magneto-elastic interaction.

\section*{Acknowledgements}

The spectroscopic work at Rutgers was supported 
by NSF Grant No. DMR-1709161. 
Sample characterization (W.Z.) was supported in part by 
the U.S. Department of Energy,
Office of Basic Energy Sciences, Division of Materials Sciences and
Engineering under Contract No. DE-SC0005463. 
The crystal growth at Rice was supported by the U.S. Department of Energy, Office
of Basic Energy Sciences, under Contract No. DE-SC0012311 and 
Robert\,A.\,Welch Foundation Grant No. C-1839.


%

\end{document}